\documentclass[a4paper,11pt]{article}
\pdfoutput=1 

\usepackage{jcappub} 

\usepackage{graphicx}
\usepackage{dcolumn}
\usepackage{amsmath}
\usepackage{multirow}

\newcommand{\Mpc}{\rm{~km~s^{-1}~Mpc^{-1}}}
\def\({\left(}
\def\){\right)}
\def\[{\left[}
\def\]{\right]}

\newcommand{\Rmnum}[1]{\uppercase\expandafter{\romannumeral #1\relax}}

\title{Using a multi-messenger and multi-wavelength observational strategy to probe the nature of dark energy through direct measurements of cosmic expansion history}

\author[a]{Jing-Zhao Qi,}
\author[a]{Shang-Jie Jin,}
\author[b,1]{Xi-Long Fan,}
\author[a]{Jing-Fei Zhang}
\author[a,1]{and Xin Zhang\note{Corresponding author.}}

\affiliation[a]{Department of Physics, College of Sciences, \& MOE Key Laboratory of Data Analytics and Optimization
for Smart Industry, Northeastern University, Shenyang 110819, China}
\affiliation[b]{School of Physics and Technology, Wuhan University, Wuhan 430072, China}

\emailAdd{qijingzhao@mail.neu.edu.cn}
\emailAdd{jinshangjie@stumail.neu.edu.cn}
\emailAdd{xilong.fan@whu.edu.cn}
\emailAdd{jfzhang@mail.neu.edu.cn}
\emailAdd{zhangxin@mail.neu.edu.cn}

\abstract{In the near future, the redshift drift observations in optical and radio bands will provide precise measurements on $H(z)$ covering the redshift ranges of $2<z<5$ and $0<z<0.3$. In addition, gravitational wave (GW) standard siren observations could make measurements on the dipole anisotropy of luminosity distance, which will also provide the $H(z)$ measurements in the redshift range of $0<z<3$. In this work, we propose a multi-messenger and multi-wavelength observational strategy to measure $H(z)$ based on the three next-generation projects, E-ELT, SKA, and DECIGO, and we wish to see whether the future $H(z)$ measurements could provide tight constraints on dark-energy parameters. The dark energy models we consider include $\Lambda$CDM, $w$CDM, CPL, HDE, and I$\Lambda$CDM models. It is found that E-ELT, SKA1, and DECIGO are highly complementary in constraining dark energy models. Although any one of these three data sets can only give rather weak constraints on each model we consider, the combination of them could significantly break the parameter degeneracies and give much tighter constraints on almost all the cosmological parameters. Moreover, we find that the combination of E-ELT, SKA1, DECIGO, and CMB could further improve the constraints on dark energy parameters, e.g., $\sigma(w_0)=0.024$ and $\sigma(w_a)=0.17$ in the CPL model, which means that these three promising probes will play a key role in helping reveal the nature of dark energy.}

\arxivnumber{2102.01292}

\begin{document}
\maketitle
\flushbottom

\section{Introduction}
One of the most important missions in modern cosmology is to understand the fundamental nature of dark energy. To achieve this goal, the basic premise is to precisely measure the equation of state (EoS) of dark energy [characterized by $w(z)$ with $z$ being redshift] through cosmological observations. However, the most difficult point for studying dark energy lies in the fact that the EoS of dark energy $w(z)$ is not a direct observable, which can only be inferred from the various actual cosmological observations, such as the measurements for the expansion history of the universe [characterized by the Hubble parameter $H(z)$]. More unfortunately, the Hubble parameter $H(z)$ at different $z$ is actually also very difficult to be measured, and instead usually the measurements of the distance--redshift relation are used to infer the EoS of dark energy.

However, in the application of distance--redshift relation measurements, there is a severe disadvantage in inferring the EoS of dark energy, i.e., lots of information is lost because of the two integrals relating cosmic distance to the EoS of dark energy. In addition, actually, it is also rather difficult to measure the absolute cosmological distances, no matter for the standard candle method [using type Ia supernovae (SNe Ia) to measure luminosity distance] and the standard ruler method [using baryon acoustic oscillation (BAO) to measure angular diameter distance].

The successful detections of gravitational waves (GWs) bring some new light to cosmological research. Through the analysis for the GW's waveform, absolute luminosity distance can be measured, which is extremely important for cosmology and referred to as {\it standard siren}. The campaign in observing the event of binary neutron star (BNS) merger (known as GW170817 \citep{TheLIGOScientific:2017qsa} as a GW event) by detecting GWs and electromagnetic waves (EMWs) in various bands opened a new era of multi-messenger astronomy \citep{GBM:2017lvd}. The observations of GWs and their EM counterparts enable us to establish an absolute luminosity distance--redshift relation, which can play an extremely important role in breaking the cosmological parameter degeneracies generated by the traditional EMW cosmological probes \citep{Bian:2021ini,Wang:2018lun,Zhang:2018byx,Zhang:2019loq,Zhang:2019ple,Wang:2019tto,Zhang:2019ylr,Li:2019ajo,Jin:2020hmc,Zhao:2019gyk,Jin:2021pcv,Wang:2021srv}. Nevertheless, even though the GW standard sirens are used, the information loss still exists in inferring the EoS of dark energy due to the two integrals in the expression of distance relating to $w(z)$. Thus, the best way of constraining $w(z)$ is to directly measure $H(z)$, instead of the distance--redshift relation. Actually, the ways of directly measuring $H(z)$ by using GW observations have been proposed.

For instance, Seto et al. \cite{seto2001possibility} proposed that using the DECihertz Interferometer Gravitational wave Observatory (DECIGO) to observe BNS merger during about a 10-year observation could measure the phase and frequency evolutions of GW signal which are proportional to $H(z)$. However, this approach strongly depends on the knowledge of waveform and is limited by the estimation error on the individual source mass. Another approach of directly measuring $H(z)$ through GW observation is to use dipole components of luminosity distance arising from the matter inhomogeneities of large-scale structure and the local motion of observer \citep{Sasaki:1987ad}. This idea was proposed by Bonvin et al. \cite{Bonvin:2005ps,Bonvin:2006en} for the SN Ia observation, and further developed by Nishizawa et al. \cite{Nishizawa:2010xx} for the GW observation.


Compared with the SN observation, the GW observation can have a larger number of samples, smaller systematic errors, and higher redshifts in the future. According to the design and objectives of DECIGO \citep{kawamura2021current}, the detectable number of GW events from BNS merger could reach $10^5$ per year within a redshift of $5$. Moreover, since the signals can be observed several years before the merge, the accuracies of the direction of GW source and the predicted coalescence time could be greatly improved, making the tracking of its electromagnetic counterpart more reliable. Therefore, in the GW multi-messenger astronomy era, we can use this method to measure $H(z)$, and then to explore the nature of dark energy.

On the other hand, actually, using EMWs can also directly measure $H(z)$. In addition to the traditional ways such as the differential age method \citep{stern2010cosmic} and radial BAO method \citep{gaztanaga2009clustering}, in the future, the most promising way is to use the redshift drift, also known as the Sandage-Loeb (SL) test, which was originally proposed by Sandage \cite{sandage1962change} and further improved by Loeb \cite{Loeb:1998bu}. By the upcoming experiments such as the European Extremely Large Telescope (E-ELT) and the Square Kilometre Array (SKA), the precise measurements of the redshift drift will be achieved through two different means. In the optical band, E-ELT with high-resolution optical spectrograph enables the measurement of redshift drift in the redshift range of $2<z<5$ \citep{Liske:2008ph,Quercellini:2010zr} by observing Lyman-$\alpha$ absorption lines of distance quasars. By observing the neutral hydrogen 21-cm emission signals of galaxies in the radio band instead of the Lyman-$\alpha$ absorption lines, the SKA1-mid can measure the redshift drift in the redshift range of $0<z<0.3$ \citep{Klockner:2015rqa,Martins:2016bbi}.

In the forthcoming 10-20 years, we will be in an era of multi-messenger and multi-wavelength astronomical observations, and the relevant projects will provide highly precise observational data. In particular, the synergies between GW and EMW observations in various bands will play a crucial role in precisely measuring the cosmological parameters, which would help answer the fundamental questions in cosmology, such as the nature of dark energy. 
In this work, we provide such an example. As mentioned above, some next-generation projects can offer direct measurements on $H(z)$, which is rather important for the studies on dark energy. We wish to investigate what the synergy between GW and EMW observations will bring to cosmology in the near future. 

On the other hand, as we know, the measurements for the cosmic microwave background (CMB) move us into the era of precision cosmology. However, since dark energy dominates the late universe, the measurements of CMB in the early universe have very weak constraints on the evolution of dark energy. It is necessary to combine with other late universe probes to break the degeneracy between cosmological parameters so that dark energy can get better constraints. Therefore, in this paper, the second point we want to investigate is what effect the combination of these three promising $H(z)$ measurements of the late universe and CMB of the early universe will have on breaking the degeneracies between cosmological parameters and what degree of constraint precision can be improved to dark energy. Here we consider five typical dark energy models, i.e., the $\Lambda$CDM model, the $w$CDM model, the Chevalliear-Polarski-Linder (CPL) model \citep{Chevallier:2000qy,Linder:2002et}, the holographic dark energy (HDE) model and a specific interacting dark energy model -- I$\Lambda$CDM, to discuss these problems we are interested in.


\section{Method and data}

\subsection{Redshift drift from optical and radio observations}

In an observing time interval ($\Delta t$), the shift in the spectroscopic velocity of a source ($\Delta v$) can be expressed as~\citep{sandage1962change,Liske:2008ph}
\begin{equation}
\Delta v=\frac{\Delta z}{1+z}=H_0\Delta t \(1-\frac{E(z)}{1+z}\),
\end{equation}
where $E(z)\equiv H(z)/H_0$ is the dimensionless Hubble expansion rate, and $H_0\equiv H(z=0)$ is the Hubble constant. We set $G=c=1$ throughout this paper.

\begin{figure*}
\centering
\includegraphics[scale=0.5]{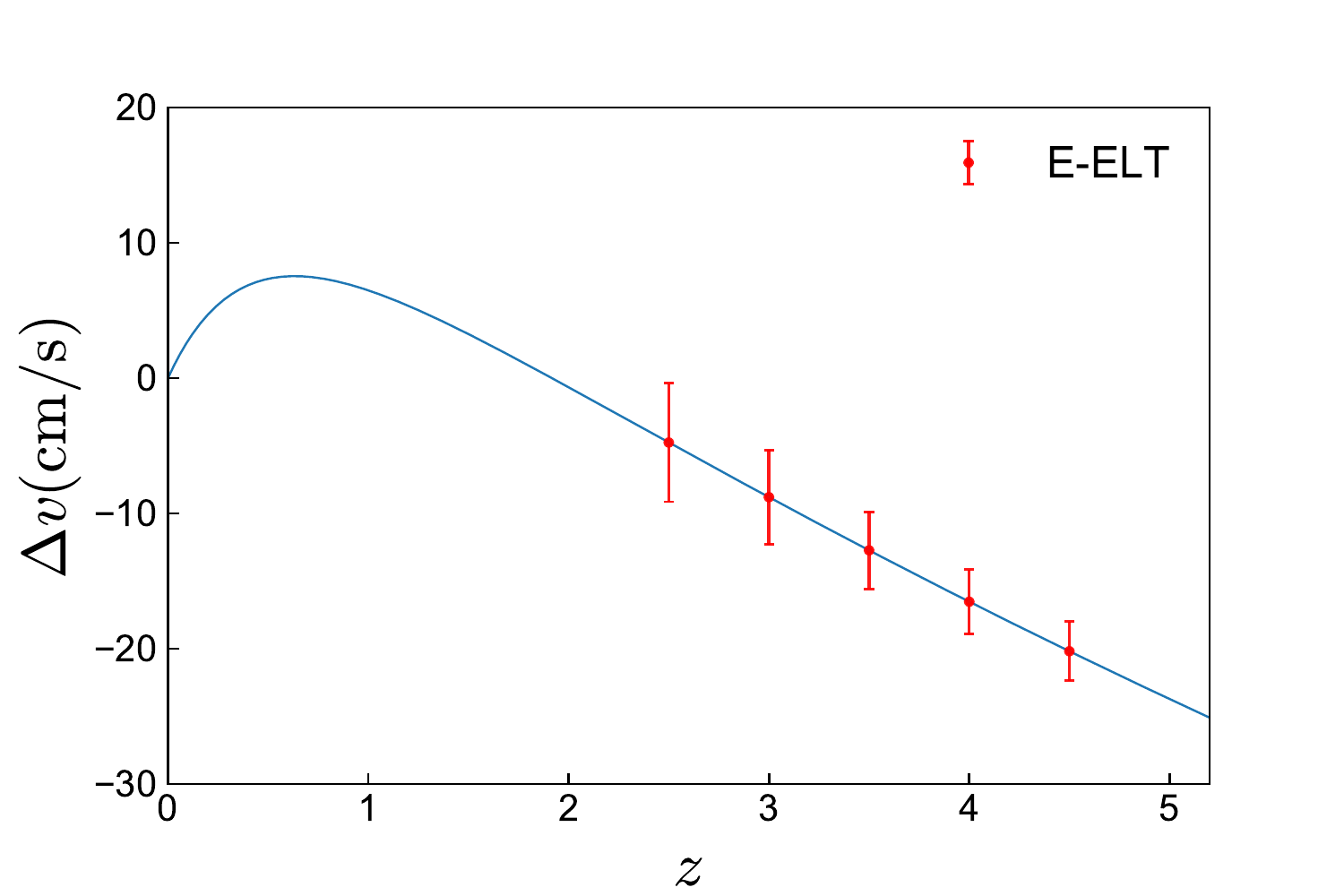}
\includegraphics[scale=0.5]{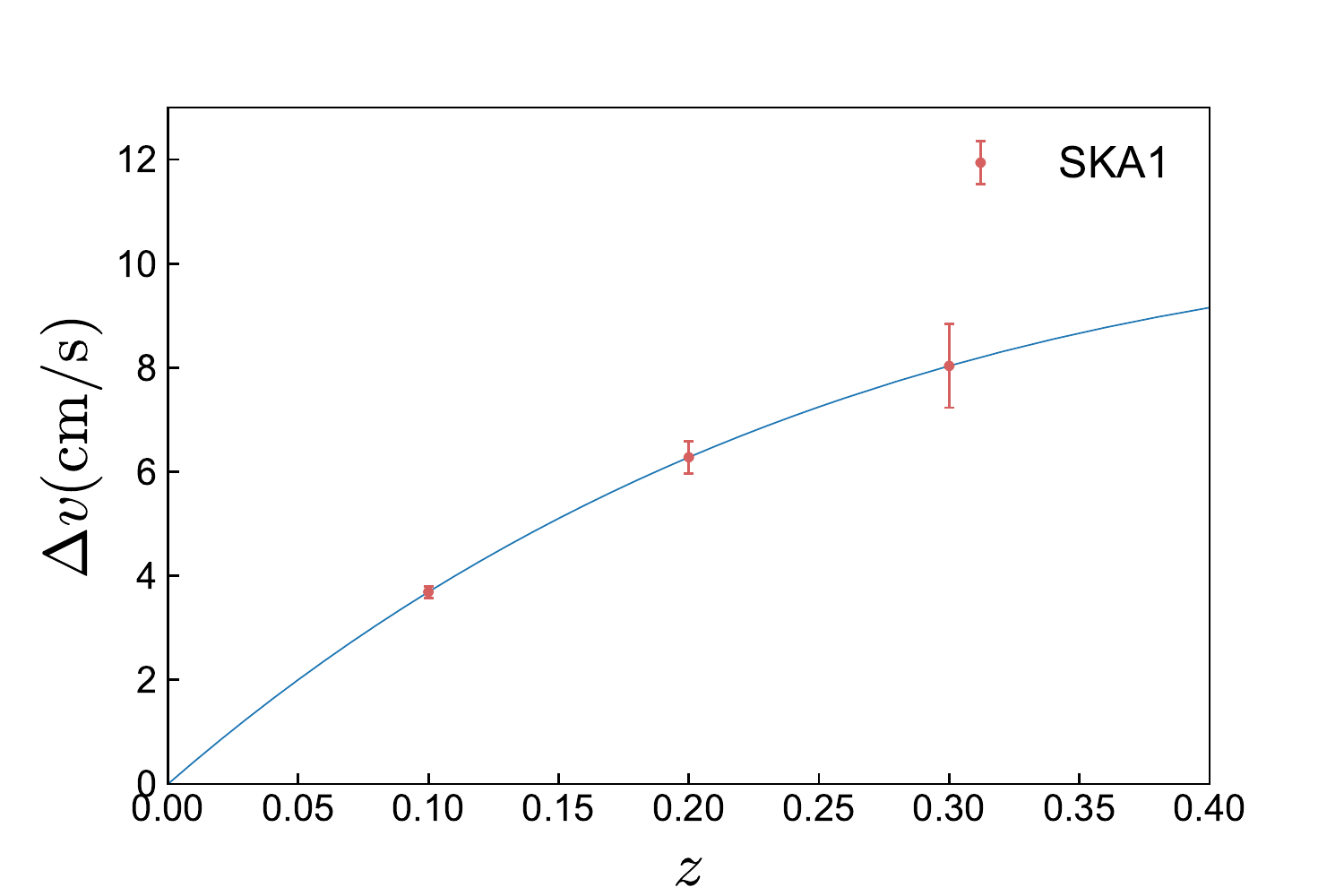}
\caption{The predicted $\Delta v$ measurements from E-ELT (left panel) and SKA1 (right panel).}\label{redshift}
\end{figure*}

By observing the Lyman-$\alpha$ absorption lines of distant quasar, E-ELT with high-resolution optical spectrograph could measure the velocity shift in a redshift range $2<z<5$ \citep{Liske:2008ph,Loeb:1998bu}. The possibility of detecting the redshift drift with E-ELT was analysed in refs.~\citep{Liske:2008ph,Quercellini:2010zr}. Based on the Monte Carlo simulations for E-ELT, the achievable precision of $\Delta v$ can be estimated as \citep{Liske:2008ph}
\begin{equation}
\sigma_{\Delta v}=1.35\(\frac{S/N}{2370}\)^{-1}\(\frac{N_{\rm{QSO}}}{30}\)^{-1/2}\(\frac{1+z_{\rm{QSO}}}{5}\)^{q}~\rm{cm~s^{-1}},
\end{equation}
with $q=-1.7$ for $z\leq 4$, and $q=-0.9$ for $z>4$, where $S/N$ is the signal-to-noise ratio of the Lyman-$\alpha$ spectrum, and $N_{\rm{QSO}}$ is the number of observed quasars at the effective redshift $z_{\rm{QSO}}$. In this work, we assume a time span of $\Delta t=30$ yr, a signal-to-noise ratio $S/N=3000$, and $N_{\rm{QSO}}=6$ quasars in each of five redshift bins at effective redshifts $z_i=[2.5, 3, 3.5, 4, 4.5]$, namely a total of 30 observable quasars \citep{Martins:2016bbi,Alves:2019hrg}. In this simulation, the fiducial cosmology we adopt is the $\Lambda$ cold dark matter ($\Lambda$CDM) model with $\Omega_m=0.315$ and $H_0=67.4 \rm{~km~s^{-1}~Mpc^{-1}}$ from the \textit{Planck} 2018 results \citep{Aghanim:2018eyx}. The simulated redshift-drift data of E-ELT are shown in the left panel of figure \ref{redshift}.

For the SKA1-mid, by observing the neutral hydrogen (HI) emission signals of galaxies instead of the Lyman-$\alpha$ absorption lines, it can measure the redshift drift in the redshift range of $0<z<0.3$ as an important supplement to E-ELT. Based on the sensitivity estimates of the SKA and the number counts of the expected HI galaxies, detailed studies of observational uncertainties for the redshift drift have been done \citep{Klockner:2015rqa,Martins:2016bbi}. Following them, for SKA1-mid in an observational timespan of 40 years, we produce 3 mock data of the drift $\Delta v$ in redshift bins centered on $z_i=[0.1,0.2,0.3]$ with uncertainties $\sigma_{\Delta v}$, respectively, of 3\%, 5\%, and 10\%, which are presented in the right panel of figure \ref{redshift}.

\subsection{Dipole anisotropy of luminosity distance from GW multi-messenger observation}

In a homogeneous and isotropic universe, in principle, the observations of luminosity distances to astronomical objects over the sky should not be directional. However, in fact, due to the matter inhomogeneities of the large-scale structure and the local motion of observer \citep{Sasaki:1987ad}, there are tiny anisotropies leading to the appearance of correction term for the luminosity distance $d_L$, given by \citep{Bonvin:2006en,Bonvin:2005ps,Nishizawa:2010xx}
\begin{eqnarray}
d_{L}(z, \mathbf{n})&=&d_{L}^{(0)}(z)+d_{L}^{(1)}(z)\cos \theta, \\
d_{L}^{(0)}(z) &\equiv & \frac{1}{4 \pi} \int d \mathbf{n} d_{L}(z, \mathbf{n}), \\
d_{L}^{(1)}(z) &\equiv & \frac{3}{4 \pi} \int d \mathbf{n}(\mathbf{n} \cdot \mathbf{e}) d_{L}(z, \mathbf{n}),
\end{eqnarray}
where $\cos \theta=\mathbf{n} \cdot \mathbf{e}$, $\mathbf{n}$ is the angular position of the luminous object, and $\mathbf{e}$ is the unit vector directed toward dipole coming from the peculiar velocity of the observer. In a homogeneous and isotropic universe, $d_{L}^{(0)}(z)$ can be reduced to the traditional meaning of luminosity distance,
\begin{equation}
d_{L}^{(0)}(z)=(1+z) \int_{0}^{z} \frac{d z^{\prime}}{H\left(z^{\prime}\right)}.
\end{equation}
The dipole $d_{L}^{(1)}(z)$ can be expressed as 
\begin{align}
d_L^{(1)}(z)= \frac{|\mathbf{v}_0|(1+z)^2}{H(z)},
\label{eq:dL1_Doppler}
\end{align}
with the direction of the dipole specifically chosen as $\mathbf{e}=\mathbf{v}_0/|\mathbf{v}_0|$, and the value of $|\mathbf{v}_0|$ is estimated as $369.1 \pm 0.9~\mathrm{km}~\mathrm{s}^{-1}$ in the CMB frame \citep{Jarosik:2010iu}. Thus, a direct measurement of $H(z)$ can be given by the dipole anisotropy of luminosity distance. The error of $H(z)$ is related to the error of the luminosity distance, and can be estimated as \citep{Bonvin:2006en,Bonvin:2005ps,Nishizawa:2010xx}
\begin{equation}
\frac{\Delta H(z)}{H(z)}=\frac{\Delta d_{L}^{(1)}(z)}{d_{L}^{(1)}(z)}=\sqrt{3}\left[\frac{d_{L}^{(1)}(z)}{d_{L}^{(0)}(z)}\right]^{-1}\left[\frac{\Delta d_{L}^{(0)}(z)}{d_{L}^{(0)}(z)}\right].
\end{equation}
For detailed description of the above formulae derivation, we refer the reader to refs.~\citep{Bonvin:2006en,Bonvin:2005ps,Nishizawa:2010xx}. In this paper, to get $\Delta d_{L}^{(0)} / d_{L}^{(0)}$, we consider the dipole anisotropy of luminosity distance from the GW standard sirens produced by the BNS and observed by the DECIGO detector. The systematic error of the averaged luminosity distance is~\citep{Nishizawa:2010xx}
\begin{equation}
\left[\frac{\Delta d_{L}^{(0)}(z)}{d_{L}^{(0)}(z)}\right]^{2}=\sigma_{\text {inst }}^{2}(z)+\sigma_{\text {lens }}^{2}(z)+\sigma_{\mathrm{pv}}^{2}(z),
\end{equation}
where $\sigma_{\text {inst }}$ is the uncertainty associated with the DECIGO described in detail in Sec.~\Rmnum{3}A of ref.~\citep{Nishizawa:2010xx}. $\sigma_{\text {lens }}$ is the lensing error of the form
\begin{equation}
\sigma_{\text {lens }}(z)=0.066\left[\frac{1-(1+z)^{-0.25}}{0.25}\right]^{1.8}.
\end{equation}
The peculiar-velocity error $\sigma_{\mathrm{pv}}$ is a kind of Doppler effect caused by the motions of galaxies, and it is estimated as~\citep{Gordon:2007zw}
\begin{equation}
\sigma_{\mathrm{pv}}(z)=\left|1-\frac{(1+z)^{2}}{H(z) d_{L}^{(0)}(z)}\right| \sigma_{\mathrm{v}, \mathrm{gal}},
\end{equation}
where $\sigma_{\mathrm{v}, \mathrm{gal}}=300 ~\mathrm{km}~\mathrm{s}^{-1}$ represents the 1-dimensional velocity dispersion of the galaxy \citep{Silberman:2001fa}.

\begin{figure*}
\centering
\includegraphics[scale=0.5]{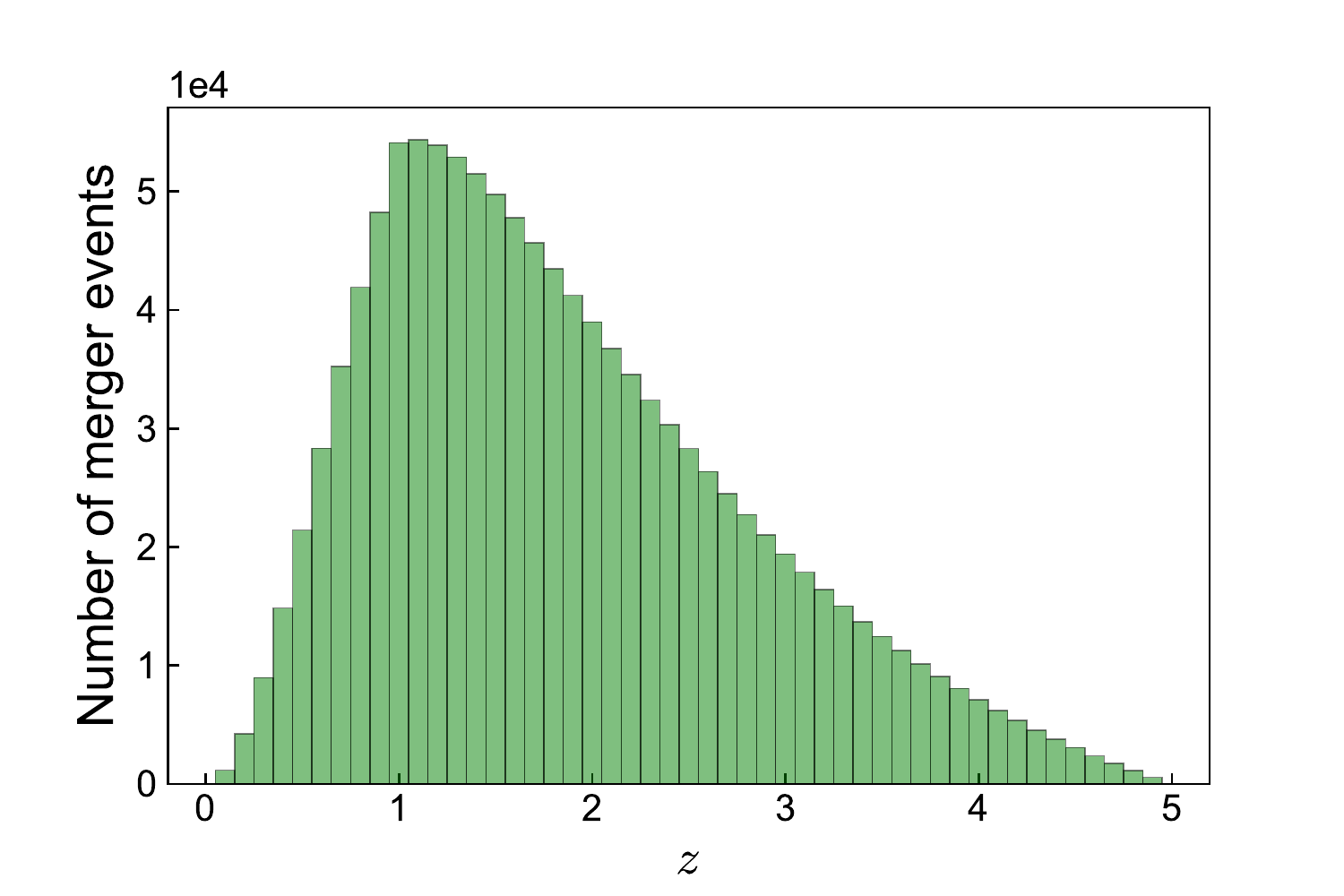}
\includegraphics[scale=0.5]{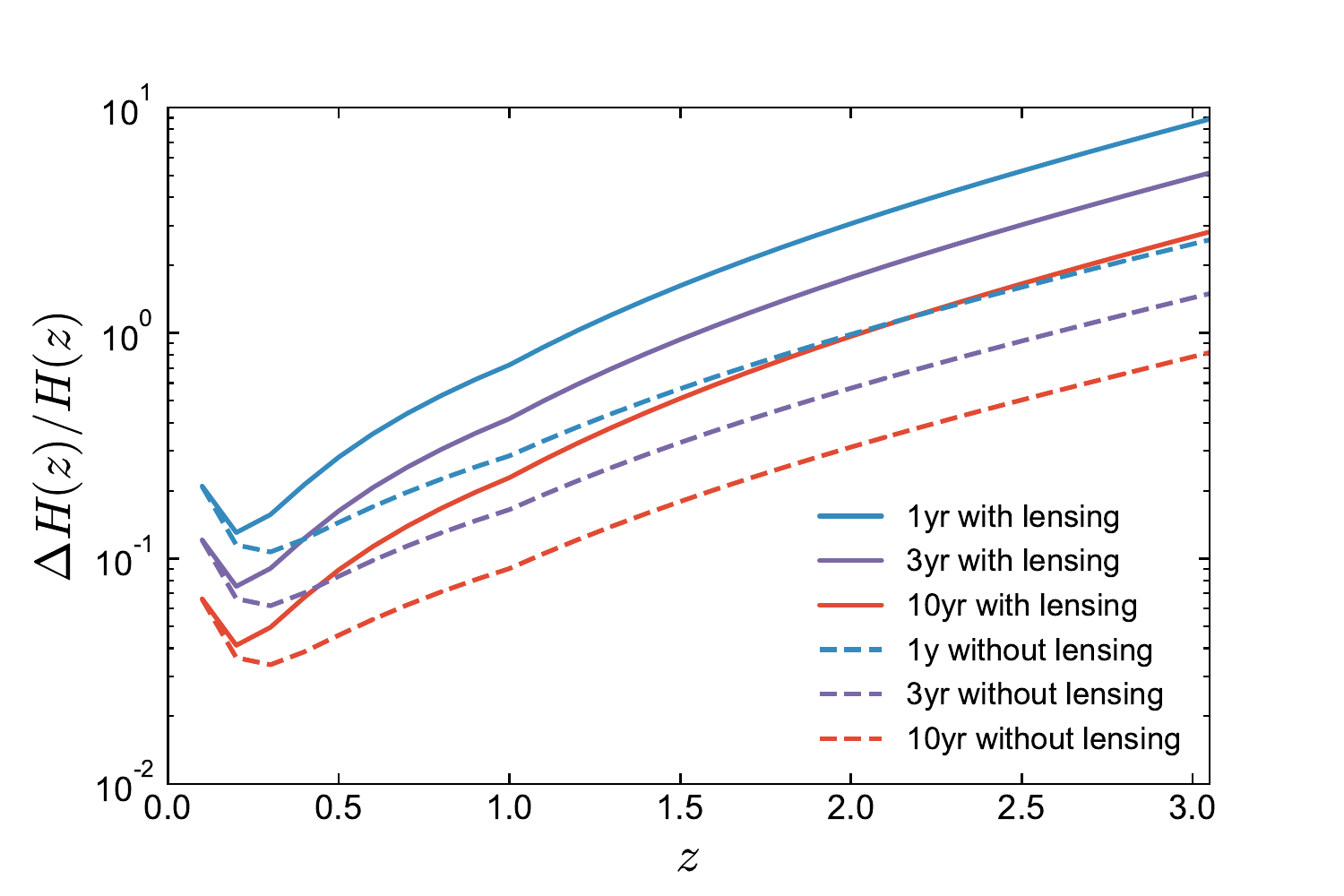}
\caption{Left panel: The distribution of events number of BNS mergers that could be detected by DECIGO during 10 years. Right panel: The relative errors of $H(z)$ for DECIGO with and without lensing errors during 1-year, 3-year, and 10-year observation time. }\label{DICIGO}
\end{figure*}

If we have $\Delta N(z)$ independent binary NS-NS systems in the vicinity of the redshift $z$, the mean error of $H(z)$ reduces to $\Delta H(z)/\sqrt{\Delta N}$, which is dependent on the number distribution of NS binary system at different redshift and the observation time. We adopt the following fitting formula of the redshift distribution of GW sources \citep{Sathyaprakash:2009xt}
\begin{equation}
P(z) \propto \frac{4 \pi d_{C}^{2}(z) R(z)}{H(z)(1+z)}, \quad R(z)=\left\{\begin{array}{ll}
1+2 z & z \leq 1 \\
\frac{3}{4}(5-z) & 1<z \leq 5 \\
0 & 5<z,
\end{array}\right.
\end{equation}
where $d_{C}(z)=\int_0^z 1/H(z) dz$ represents the comoving distance. This form of the redshift distribution of BNS mergers is obtained according to cosmic star formation rate \citep{Cutler:2009qv,Schneider:2000sg}, and it is also the commonly adopted form in the simulation of GW standard sirens. We simulate $10^6$ GW events from BNS mergers as the expectation of DECIGO in its 10-year observation \citep{kawamura2021current}. In the left panel of figure \ref{DICIGO}, we plot the number of observable events with redshift width $\Delta z=0.1$.

For the uncertainty of standard sirens, the lensing effect is a crucial source. However, in refs \citep{Nishizawa:2010xx,Zhang:2019mdf}, the $\sigma_{\rm{lens}}$ was ignored due to the authors believe the lensing systematics could be removed with some feasible techniques \citep{Hilbert:2010am,Hirata:2010ba} in the near future. For comparison, in the right panel of figure \ref{DICIGO}, the relative errors of $H(z)$ for DECIGO as a function of redshift are plotted, taking account of the case with lensing errors and the case without lensing errors, during 1-yr, 3-yr, and 10-yr observation time. In this paper, to be conservative, we will not ignore the uncertainty from the lensing effect. In addition, since the error of $H(z)$ in the redshift range of $z>3$ is too large to effectively constrain the cosmological parameters, we only select the data with the range of $0<z<3$.

\subsection{The CMB data}
For the CMB data, we use the Planck distance priors $(R,~\ell_A,~\Omega_bh^2)$ obtained from the \textit{Planck} 2018 TT,TE,EE+lowE data \citep{Aghanim:2018eyx,Chen:2018dbv}. The acoustic scale $l_A$ is defined as
\begin{equation}
\ell_{A}=\left(1+z_{*}\right) \frac{\pi D_{A}\left(z_{*}\right)}{r_{s}\left(z_{*}\right)},
\end{equation}
where $z_*$ is the redshift at the photon decoupling epoch and its approximate formula is 
\begin{equation}
z_{*}=1048\left[1+0.00124\left(\Omega_{b} h^{2}\right)^{-0.738}\right]\left[1+g_{1}\left(\Omega_{m} h^{2}\right)^{g_{2}}\right],
\end{equation}
with $g_{1}=\frac{0.0738\left(\Omega_{b} h^{2}\right)^{-0.238}}{1+39.5\left(\Omega_{b} h^{2}\right)^{0.763}}$ and $g_{2}=\frac{0.560}{1+21.1\left(\Omega_{b} h^{2}\right)^{1.81}}$ \citep{Hu:1995en}. $r_s$ is the comoving sound horizon, which is defined as
\begin{equation}
r_{s}(z)=\frac{1}{H_{0}} \int_{0}^{1 /(1+z)} \frac{d a}{a^{2} E(a) \sqrt{3\left(1+\frac{3 \Omega_{b} h^{2}}{4 \Omega_{\gamma} h^{2}} a\right)}},
\end{equation}
where $\frac{3}{4 \Omega_{\gamma} h^{2}}=31500\left(T_{\mathrm{CMB}} / 2.7 K\right)^{-4}$, and $T_{\mathrm{CMB}}=2.7255 K$.

The expression of angular diameter distance $D_A$ in a flat universe is
\begin{equation}
D_{A}(z)=\frac{1}{(1+z)H_0} \int_{0}^{z} \frac{d z^{\prime}}{E\left(z^{\prime}\right)}.
\end{equation}
The shift parameter $R$ is defined by
\begin{equation}
R\left(z_{*}\right) \equiv \left(1+z_{*}\right) D_{A}\left(z_{*}\right) \sqrt{\Omega_{m} H_{0}^{2}}.
\end{equation}
Then, $\chi^2$ of the CMB distance priors could be written as
\begin{equation}
\chi_{\mathrm{CMB}}^{2}=\left(x_{i, \mathrm{CMB}}^{\mathrm{th}}-x_{i, \mathrm{CMB}}^{\mathrm{obs}}\right)\left(C_{\mathrm{CMB}}^{-1}\right)_{i j}\left(x_{j, \mathrm{CMB}}^{\mathrm{th}}-x_{j, \mathrm{CMB}}^{\mathrm{obs}}\right),
\end{equation}
where $x_{i, \mathrm{CMB}} \equiv\left(R\left(z_{*}\right), \ell_{A}\left(z_{*}\right), \Omega_{b}h^2\right)$ with  $R=1.7502$, $\ell_A=301.471$ and $\Omega_bh^2=0.02236$ \citep{Chen:2018dbv}, and their inverse covariance matrix $C_{\mathrm{CMB}}^{-1}$ can be given as
\begin{equation}
C_{\mathrm{CMB}}^{-1}=\left(\begin{array}{ccc}
94392.3971&-1360.4913&1664517.2916 \\
-1360.4913&161.4349& 3671.6180 \\
1664517.2916& 3671.6180& 79719182.5162
\end{array}\right).
\end{equation}

\subsection{Cosmological models}
In this subsection, we briefly introduce the dark energy models including $\Lambda$CDM model, $w$CDM model, CPL model, HDE model, and I$\Lambda$CDM model.
\begin{itemize}
\item $\Lambda$CDM model with a equations of state (EoS) of the vacuum energy $w=-1$ is the most promising candidate of dark energy. Its normalized Hubble parameter defined as $E(z)\equiv H(z)/H_0$ is

\begin{equation}
E(z)=\sqrt{\Omega_{\mathrm{m}}(1+z)^{3}+\Omega_{\mathrm{r}}(1+z)^{4}+\left(1-\Omega_{\mathrm{m}}-\Omega_{\mathrm{r}}\right)}.
\end{equation}

\item $w$CDM model with a constant EoS $w$ is the simplest case for dynamical dark energy and its normalized Hubble parameter is given by
\begin{equation}
E(z)= \sqrt{\Omega_{m}(1+z)^{3}+\Omega_{r}(1+z)^{4}+\left(1-\Omega_{m}-\Omega_{r}\right)(1+z)^{3(1+w)}}.
\end{equation}

\item CPL model is a phenomenological model to explore the evolution of $w$. For this model, the forms of $w(z)$ and normalized Hubble parameter are written as, respectively,
\begin{equation}
w(z)=w_0+w_a\frac{z}{1+z},
\end{equation}
\begin{equation}
\begin{aligned}
E(z)&=&\sqrt{ \Omega_{m}(1+z)^{3}+\Omega_{r}(1+z)^{4} 
+\left(1-\Omega_{m}-\Omega_{r}\right)(1+z)^{3\left(1+w_{0}+w_{a}\right)} \exp \left(-\frac{3 w_{a} z}{1+z}\right)}.
\end{aligned}
\end{equation}

\item The holographic dark energy model is constructed by considering the holographic principle of quantum gravity theory in a quantum effective field theory \citep{Li:2004rb}. HDE model could effectively alleviate the fine-tuning and coincidence problems that are serious problems in the $\Lambda$CDM model \citep{Li:2004rb}. In this model, the density of dark energy is given by 
\begin{equation}
\rho_{\mathrm{de}} \propto M_{\mathrm{pl}}^{2} L^{-2},
\end{equation}
where $M_{\mathrm{pl}}$ is the reduced Planck mass, and $L$ is the infrared (IR) cutoff length scale in the effective quantum field theory. Different holographic dark energy models depend on the choice of different IR cutoff $L$ \citep{Zhang:2005hs,Zhao:2017urm,Li:2017usw,Zhang:2019ple,Xu:2016grp}. In this paper, we choose the event horizon size of the universe as the IR cutoff \citep{Li:2004rb}. In this case, the dark energy density is given by
\begin{equation}
\rho_{\mathrm{de}}=3 c^2 M_{\mathrm{pl}}^{2} R_{\mathrm{eh}}^{-2},
\end{equation}
where $c$ is a dimensionless phenomenological parameter determining properties of the HDE model, and it should not be confused with the speed of light.
$R_{\mathrm{eh}}$ is the event horizon size defined as
\begin{equation}
R_{\mathrm{eh}}(t)=a(t) \int_{t}^{\infty} \frac{d t^{\prime}}{a\left(t^{\prime}\right)}.
\end{equation}

The evolution of the HDE obeys the following differential equations
\begin{equation}
\begin{aligned}
&\frac{1}{E(z)} \frac{d E(z)}{d z}=-\frac{\Omega_{\mathrm{de}}(z)}{1+z}\left(\frac{1}{2}+\frac{\sqrt{\Omega_{\mathrm{de}}(z)}}{c}-\frac{3}{2 \Omega_{\mathrm{de}}(z)}\right), \\
&\frac{d \Omega_{\mathrm{de}}(z)}{d z}=-\frac{2 \Omega_{\mathrm{de}}(z)\left(1-\Omega_{\mathrm{de}}(z)\right)}{1+z}\left(\frac{1}{2}+\frac{\sqrt{\Omega_{\mathrm{de}}(z)}}{c}\right).
\end{aligned}
\end{equation}

\item Assuming some coupling between dark energy and cold dark matter, the interacting dark energy models were proposed, in which the energy conservation equations for dark energy and cold dark matter usually satisfy

\begin{eqnarray}
\dot{\rho}_{\mathrm{de}}&=&-3 H(1+w) \rho_{\mathrm{de}}+Q,\label{DE1} \\
\dot{\rho}_{\mathrm{c}}&=&-3 H \rho_{\mathrm{c}}-Q, \label{DE2}
\end{eqnarray}
where ${\rho}_{\mathrm{de}}$ and ${\rho}_{\mathrm{c}}$ are the energy densities of dark energy and cold dark matter, respectively, and a dot represents the derivative with respect to the cosmic time. For the energy transfer rate $Q$, its specific form is still an open question and many forms have been discussed in previous works \citep{Xia:2016vnp,Li:2011ga,Li:2013bya,Li:2014cee,Li:2014eha,Zhang:2017ize,Yin:2015pqa,Wang:2014oga,Guo:2017hea,Feng:2019mym,Li:2018ydj,Li:2019ajo,Li:2020gtk}. In this paper, we consider a phenomenological form of $Q=\beta H\rho_{\rm{de}}$, where $\beta$ is a dimensionless coupling parameter. Thus, once the specific form of $Q$ is determined, the expression of $E^2(z)$ can be obtained by solving the differential equations (\ref{DE1}) and (\ref{DE2}), 
\begin{equation}
\begin{aligned}
E^{2}(z)=& \Omega_{\operatorname{de}}\left(\frac{\beta}{w+\beta}(1+z)^{3}+\frac{w}{w+\beta}(1+z)^{3(1+w+\beta)}\right) \\
&+\Omega_{m}(1+z)^{3}+\Omega_{r}(1+z)^{4},
\end{aligned}
\end{equation}
where $\Omega_{m}=\Omega_{c}+\Omega_{b}$. Here, we just consider the case of I$\Lambda$CDM model corresponding to $w=-1$.
\end{itemize}
\section{Results and discussion}

\begin{figure}
\centering
\includegraphics[scale=0.35]{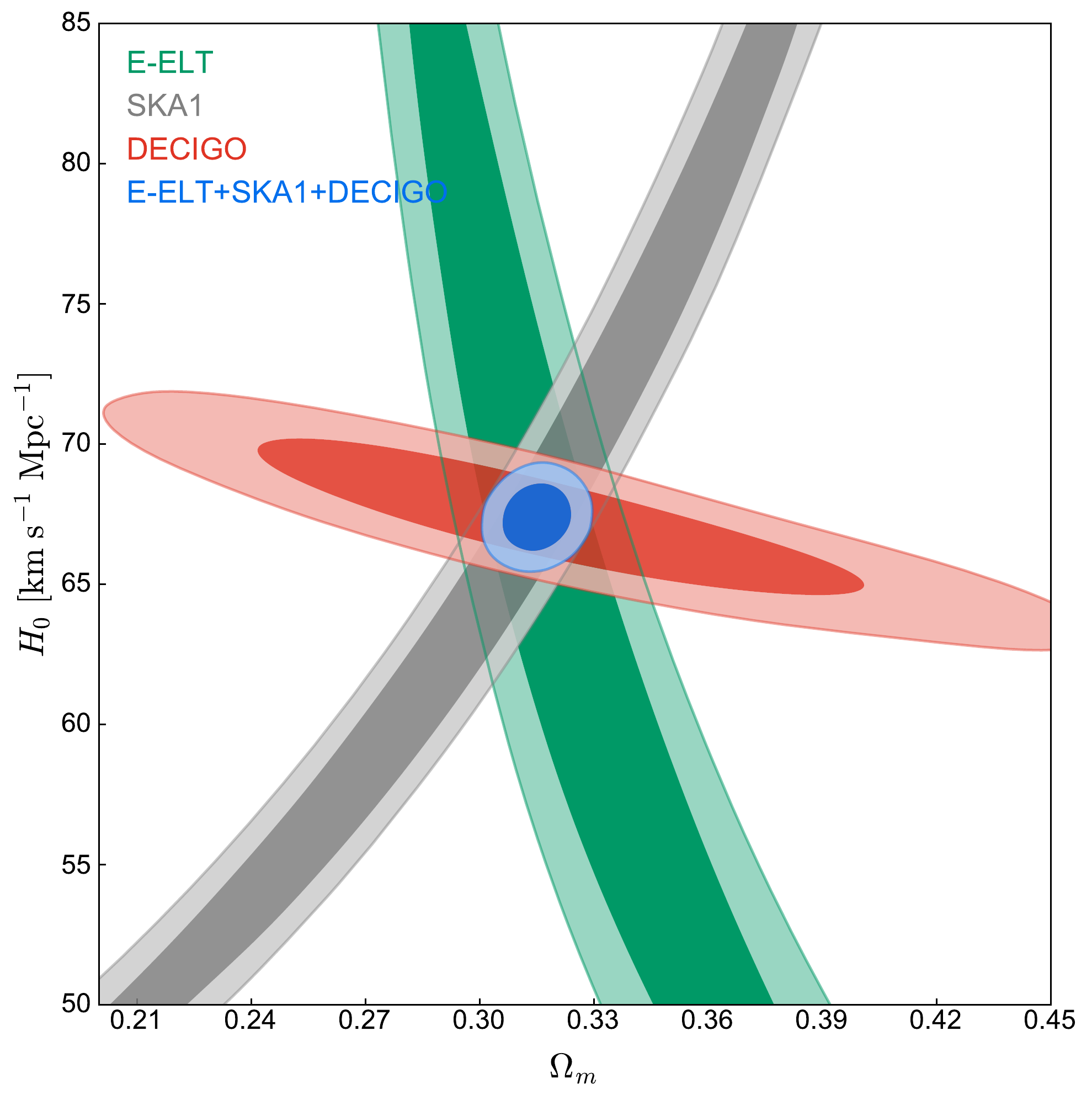}
\includegraphics[scale=0.35]{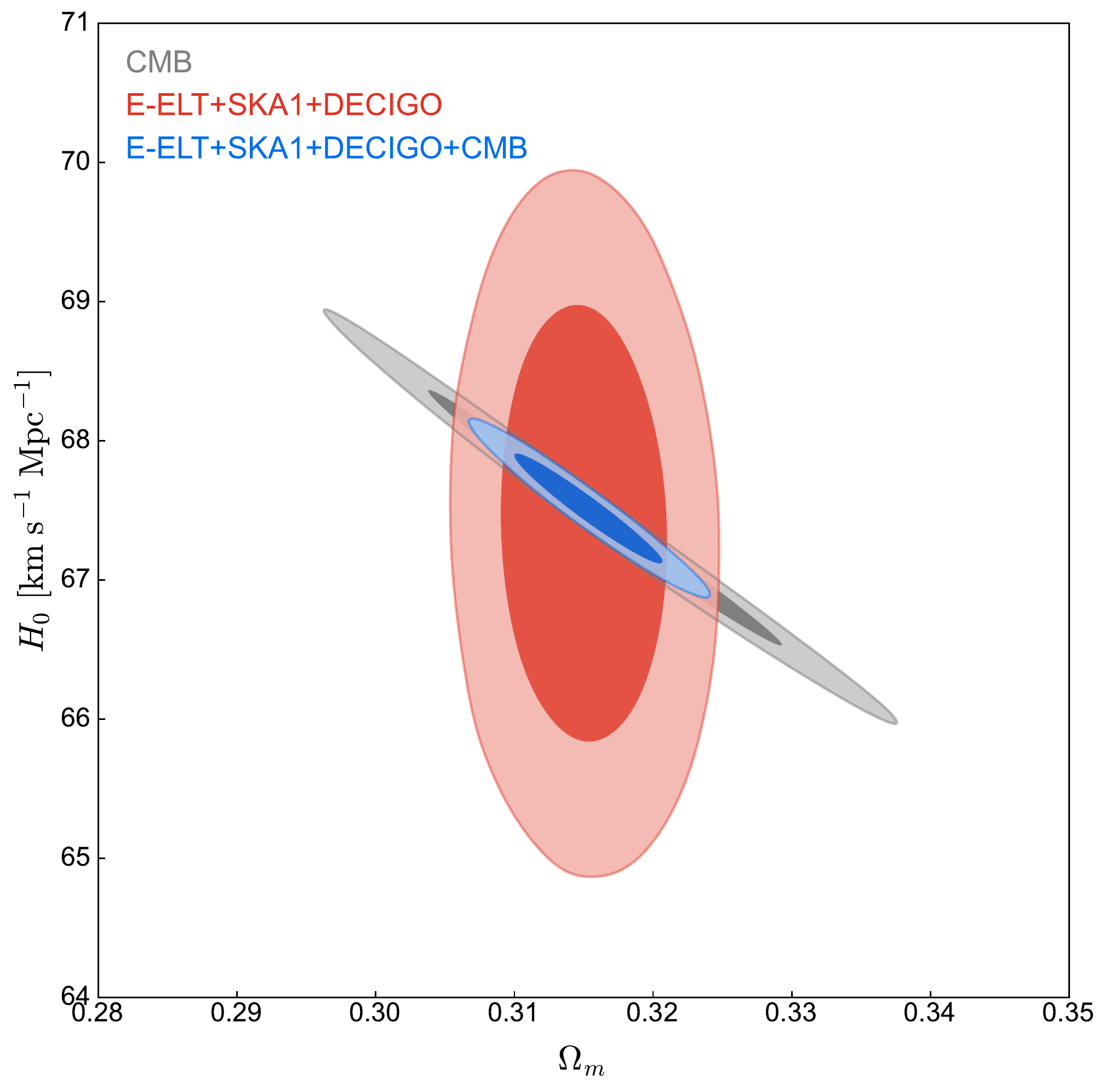}
\caption{Left panel: Constraints (68.3\% and 95.4\% confidence level) on the $\Lambda$CDM model from E-ELT, SKA1, DECIGO, and E-ELT+SKA1+DECIGO. Right panel: The confidence contours constrained from CMB, E-ELT+SKA1+DECIGO, and E-ELT+SKA1+DECIGO+CMB.}\label{Fig_LCDM}
\end{figure}

\begin{figure}
\centering
\includegraphics[scale=0.3]{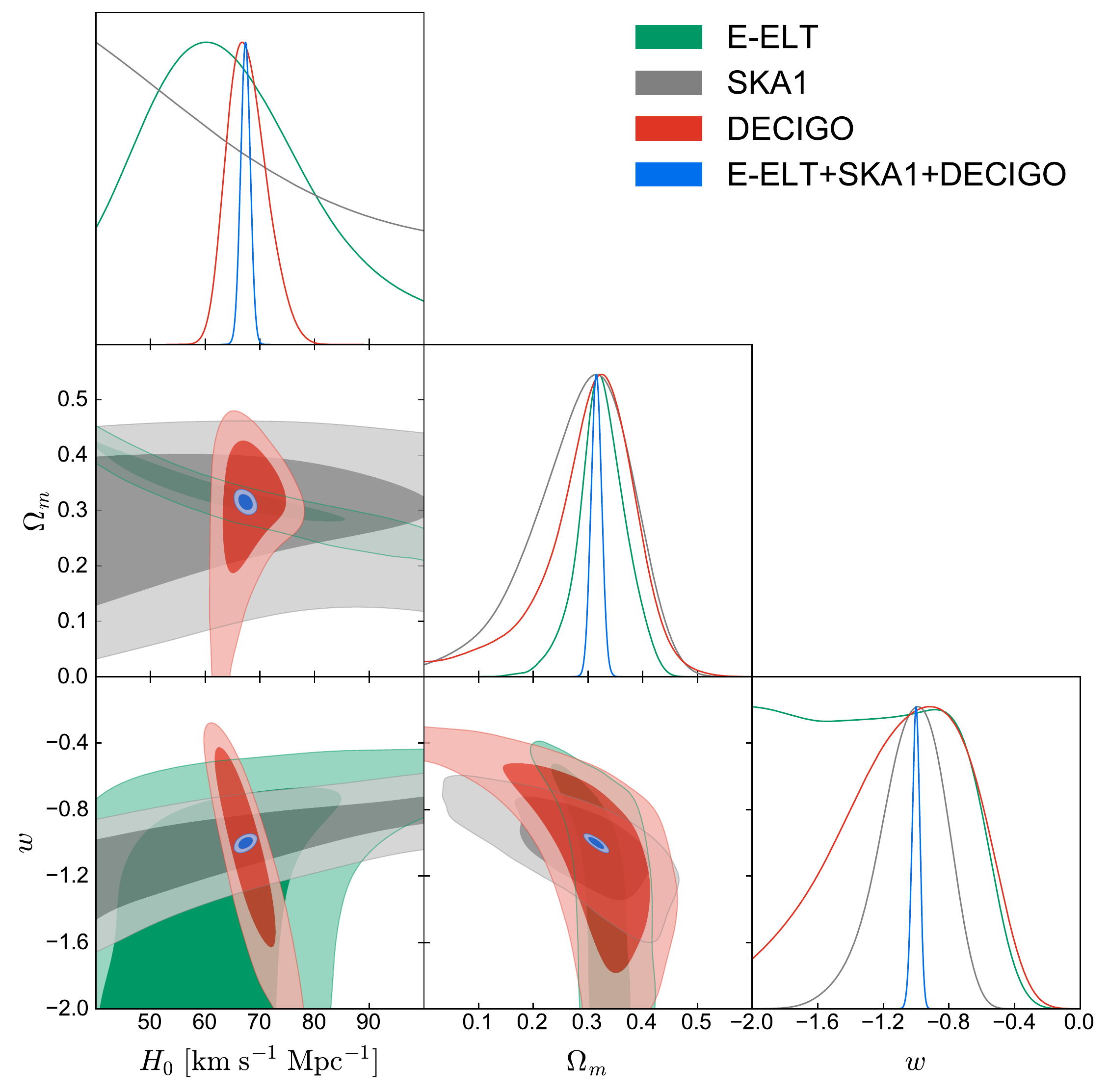}
\includegraphics[scale=0.3]{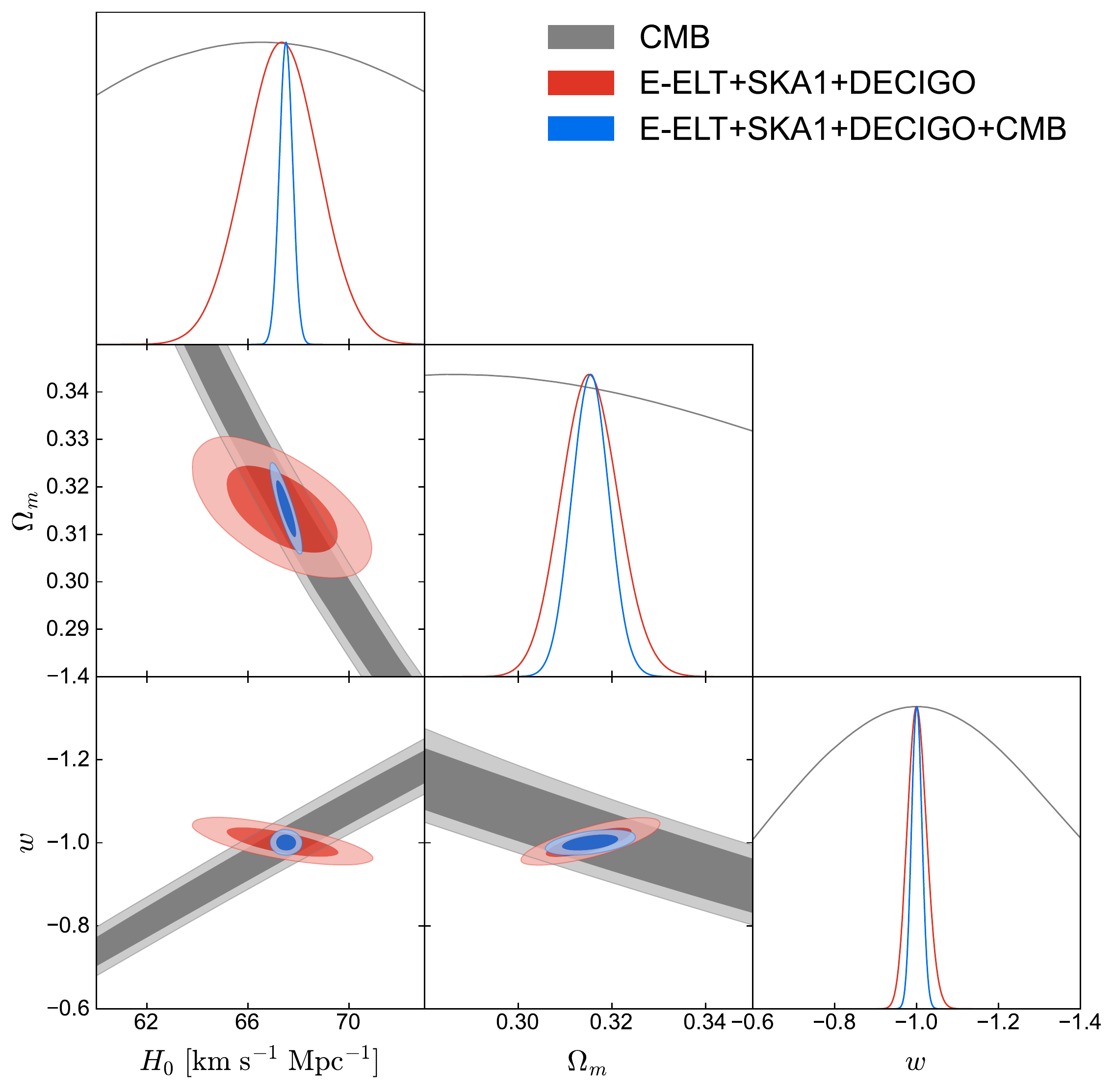}
\caption{Same as figure \ref{Fig_LCDM} but for the $w$CDM model.}\label{Fig_wCDM}
\end{figure}

\begin{figure}
\centering
\includegraphics[scale=0.3]{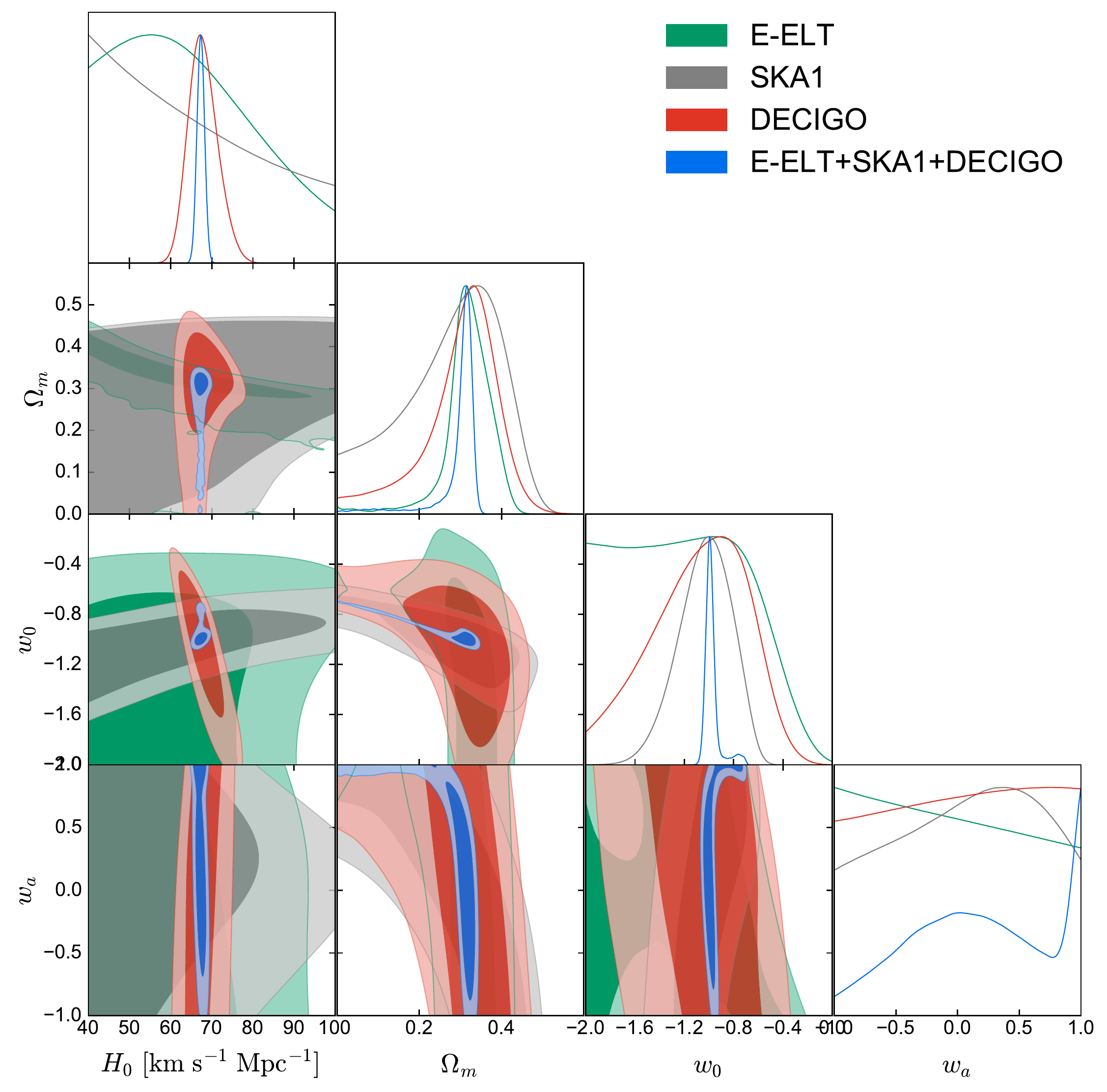}
\includegraphics[scale=0.3]{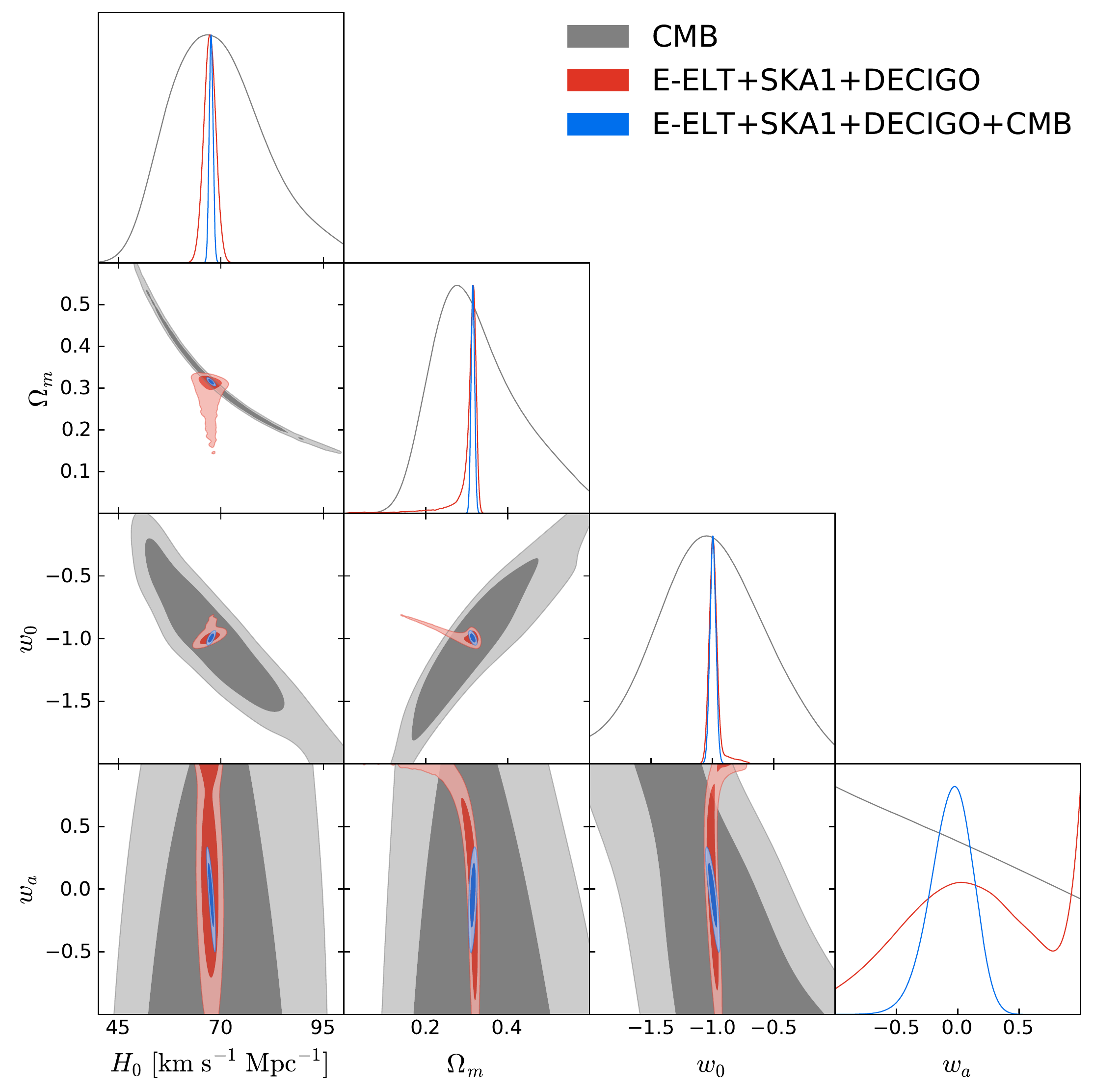}
\caption{Same as figure \ref{Fig_LCDM} but for the CPL model.}\label{Fig_CPL}
\end{figure}

\begin{figure}
\centering
\includegraphics[scale=0.3]{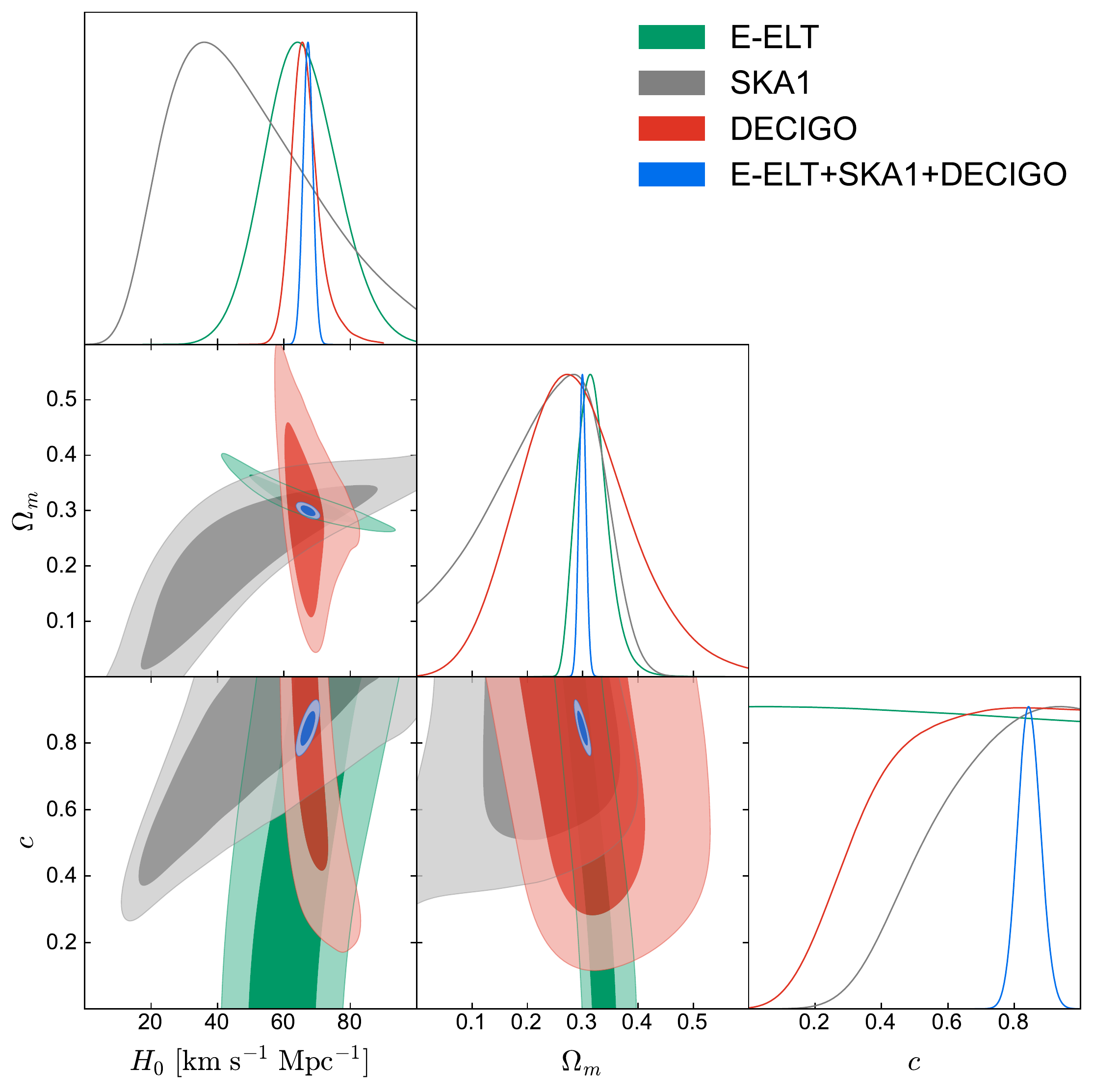}
\includegraphics[scale=0.3]{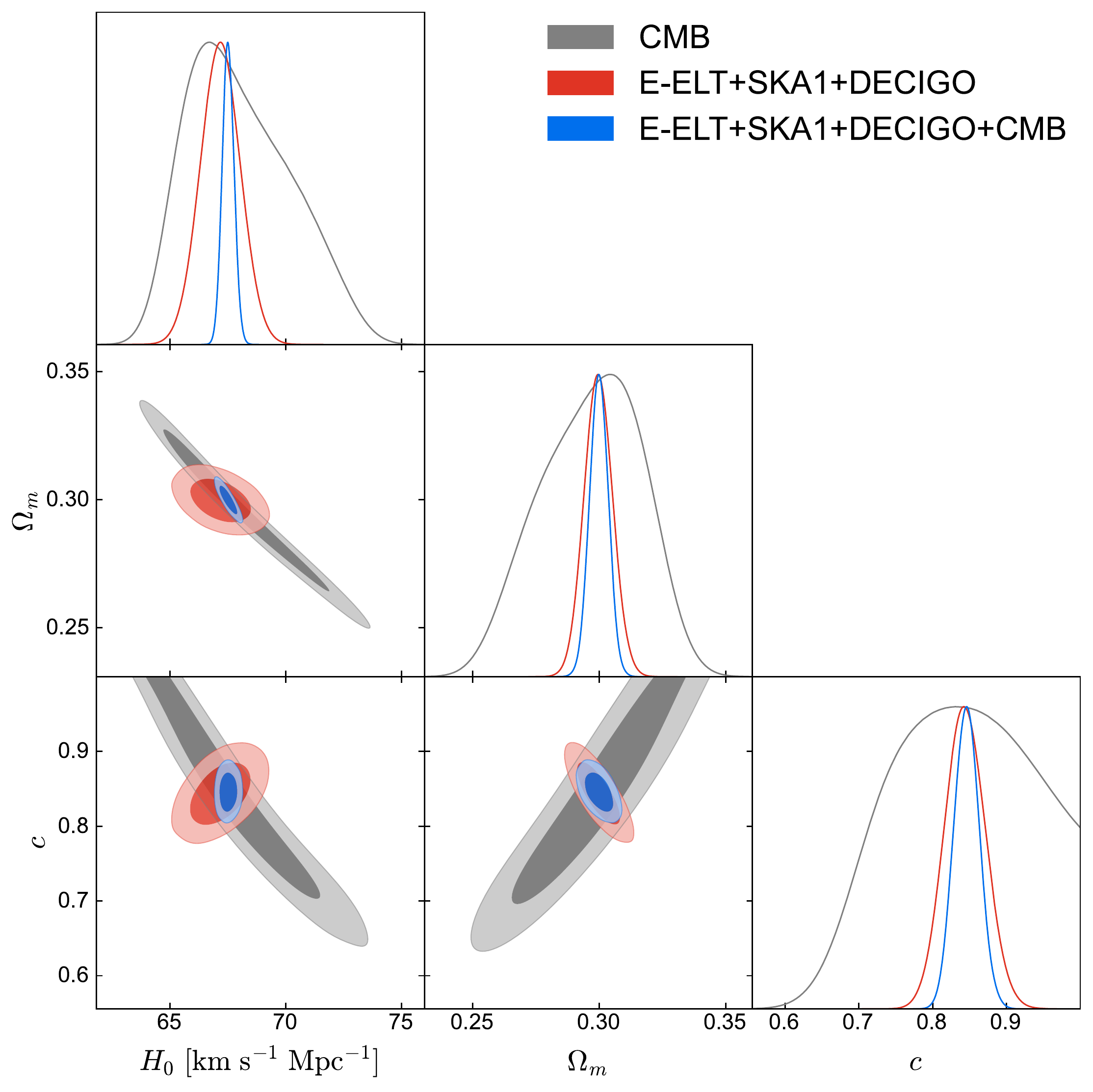}
\caption{Same as figure \ref{Fig_LCDM} but for the HDE model.}\label{Fig_HDE}
\end{figure}

\begin{figure}
\centering
\includegraphics[scale=0.3]{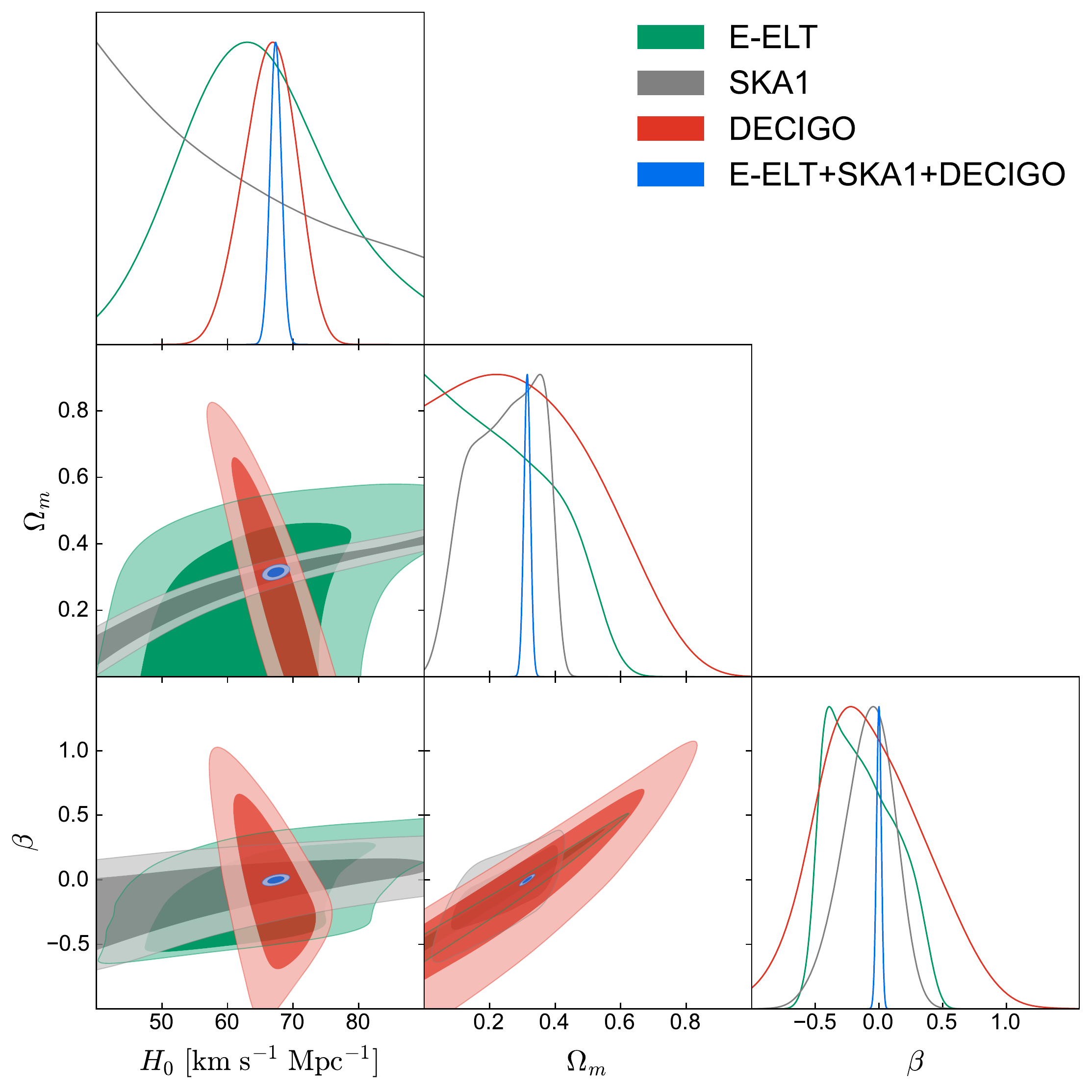}
\includegraphics[scale=0.3]{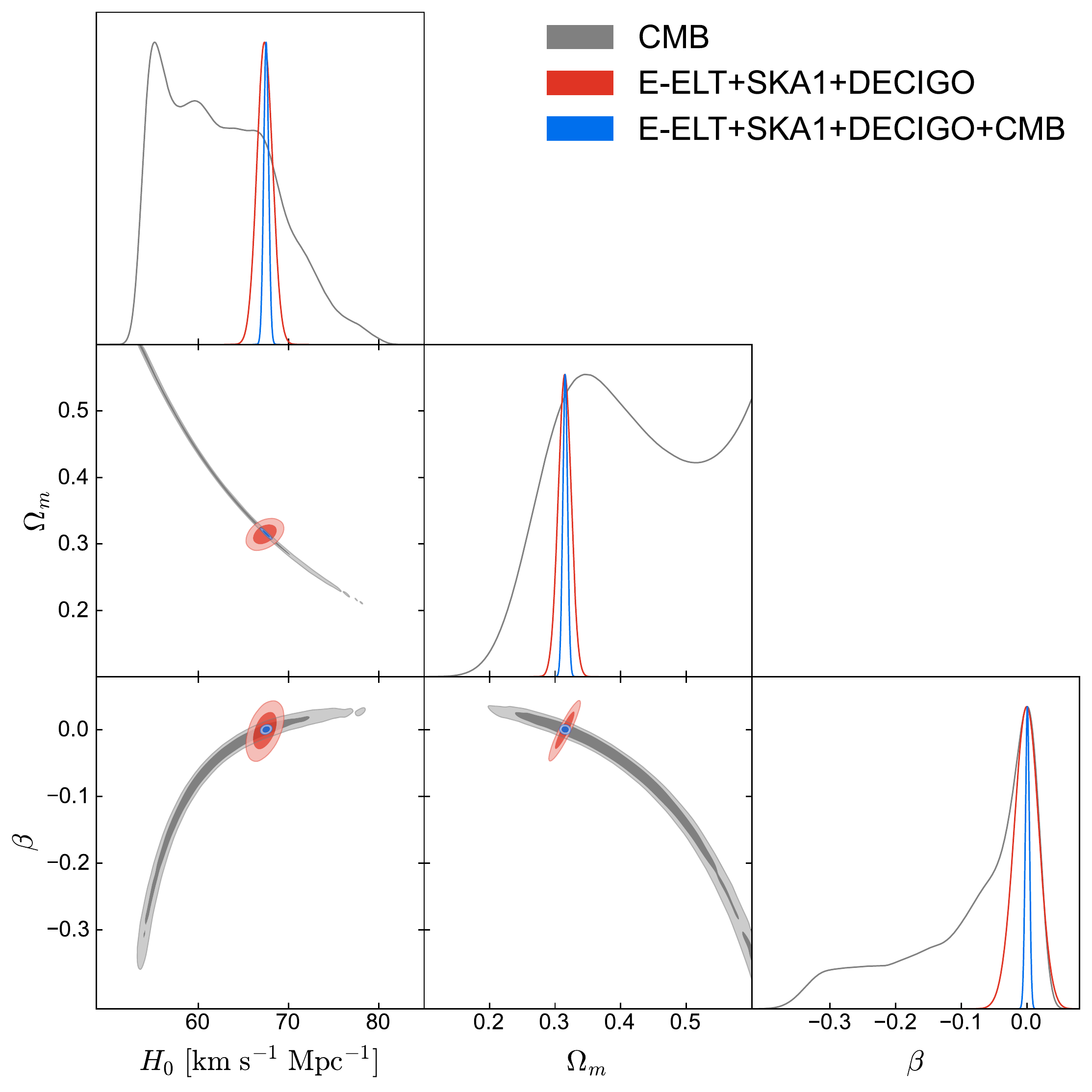}
\caption{Same as figure \ref{Fig_LCDM} but for the ILCDM model.}\label{Fig_ILCDM}
\end{figure}

\begin{table*}[!htbp]
\label{tab:full}
\setlength{\tabcolsep}{2mm}
\renewcommand{\arraystretch}{1.5}
\begin{center}{\centerline{
\begin{tabular}{|ccm{2cm}<{\centering}m{2cm}<{\centering}m{2cm}<{\centering}m{2cm}<{\centering}m{2cm}<{\centering}m{2cm}<{\centering}|}
\hline
          Model  & Error&E-ELT & SKA1 & DECIGO & ESD &CMB & ESDC \\ \hline
\multirow{2}{*}{$\Lambda$CDM}
 &$\sigma(\Omega_m)$&$0.053$  & $0.082$ & $0.052$& $0.006$ &$0.007$ & $0.0035$ \\
 &$\sigma(H_0)$&$20$ &$20$ & $1.9$ &$0.78$& $0.5$ & $0.26$ \\ \hline
 \multirow{3}{*}{$w$CDM}
 &$\sigma(\Omega_m)$&$0.042$  & $0.085$ & $0.076$ & $0.009$& $0.094$ & $0.0039$\\
 &$\sigma(H_0)$&$15$ &$-$ & $3.5$&$0.87$ &$9.5$& $0.26$ \\
 &$\sigma(w)$&$-$ &$0.20$ & $0.395$&$0.024$& $0.34$ & $0.013$ \\ \hline
 \multirow{4}{*}{CPL}
 &$\sigma(\Omega_m)$&$0.0685$  & $0.115$ & $0.0755$& $0.014$& $0.104$ & $0.0043$ \\
 &$\sigma(H_0)$&$20$&$-$ & $3.5$ &$0.91$ & $9.5$ & $0.46$ \\
 &$\sigma(w_0)$&$0.545$ &$0.215$ & $0.385$ &$0.0315$ &$0.42$ & $0.024$ \\
 &$\sigma(w_a)$&$-$ &$-$ & $-$&$-$ & $-$ & $0.17$ \\  \hline
 \multirow{3}{*}{HDE}
 &$\sigma(\Omega_m)$&$0.0275$  & $0.097$ & $0.0975$  & $0.0063$& $0.021$ & $0.0037$\\
 &$\sigma(H_0)$&$10$&$20$ & $4$  & $1.5$ & $2.3$& $0.25$\\
 &$\sigma(c)$&$-$ &$-$ & $-$  & $0.035$ & $0.101$& $0.017$\\  \hline
 \multirow{3}{*}{I$\Lambda$CDM}
 &$\sigma(\Omega_m)$&$-$  & $0.109$ & $-$ & $0.01$ & $-$&  $0.0036$\\
 &$\sigma(H_0)$&$10$&$-$ & $3.95$& $0.86$ & $6.15$& $0.27$ \\
 &$\sigma(\beta)$&$0.27$ &$0.2$ & $0.42$& $0.019$& $0.071$ & $0.0031$ \\ \hline
\end{tabular}}}
\end{center}
\caption{Constraint errors ($\sigma$) of cosmological parameters in the $\Lambda$CDM, $w$CDM, CPL, HDE, I$\Lambda$CDM models using the E-ELT, SKA1, DECIGO, ESD, CMB and ESDC data.}
\end{table*}

The redshift drift measurements from optical-band and radio-band observations and the measurements for the dipole anisotropy of luminosity distance from GW multi-messenger observations all could offer direct measurements on $H(z)$, which is important for the cosmological research. We wish to see how these $H(z)$ data could put constraints on the dark-energy parameters. The main results are given in figures \ref{Fig_LCDM}--\ref{Fig_ILCDM}. The key issue we wish to investigate is the constraint ability of these three $H(z)$ probes for dark energy, which could be presented directly by the constraint error $\sigma$ of parameters. The complete constraints results are shown in table \ref{tab:full}.

In figure \ref{Fig_LCDM}, we show the constraints on the $\Lambda$CDM model in the $\Omega_m$--$H_0$ plane from E-ELT, SKA1, DECIGO, CMB and the combinations of them, i.e., E-ELT+SKA1+DECIGO (ESD) and E-ELT+SKA1+DECIGO+CMB (ESDC). In the left panel, we can clearly see that the constraints from the redshift drift measurements by both E-ELT and SKA1 are rather weak, but they are highly complementary with each other (see also ref.~\citep{Liu:2019asq}), and the combination of them could effectively break the parameter degeneracy. The GW multi-messenger observation from DECIGO could provide a much better constraint. We can see that the dipole anisotropy of GW standard sirens could give a tight constraint on $H_0$, $\sigma(H_0)\approx 1.9 \Mpc$, although it is still much weaker than the {\it Planck} 2018 TT,TE,EE+lowE+lensing result of $\sigma(H_0)\approx 0.5\Mpc$ \citep{Aghanim:2018eyx}. Since the parameter degeneracy direction by DECIGO is different from those by both E-ELT and SKA1, the combination of them could give much tighter constraints. We find that the joint constraint on the $\Lambda$CDM model gives the results $\sigma(H_0)\approx 0.78 \Mpc$ and $\sigma(\Omega_m)\approx 0.006$, which are comparable with the {\it Planck} 2018 results \citep{Aghanim:2018eyx} [$\sigma(H_0)=0.5\Mpc$ and $\sigma(\Omega_m)=0.007$] as shown in the right panel of figure \ref{Fig_LCDM}. Moreover, when these three probes are combined with CMB data, the degeneracies between parameters could be significantly broken and then one can get more precise constraints: $\sigma(\Omega_m)=0.0035$ and $\sigma(H_0)=0.26\Mpc$.

In figure \ref{Fig_wCDM}, we show the case for the $w$CDM model (where $w$ is assumed to be a constant). In the left panel, we find that the $H(z)$ data from each one of these three observations can only give a weak constraint on this model. However, the parameter degeneracy directions of them are rather different. Specifically, it is found that E-ELT, SKA1, and DECIGO could offer better constraints on $\Omega_m$, $w$, and $H_0$, respectively, and thus the combination of them could give much tighter constraints on all parameters. In particular, we find that the joint constraint from ESD gives $\sigma(w)\approx 0.024$, which is better than the result of {\it Planck} 2018 TT,TE,EE+lowE+lensing+SNe+BAO, $\sigma(w)\approx 0.03$ \citep{Aghanim:2018eyx}. In the right panel of figure \ref{Fig_wCDM}, the comparison of CMB and ESD shows that the CMB of the early universe have a very weak constraint on the dark energy, while the combination of three directly measured $H(z)$ data could give a much tighter constraint on $w$CDM model. Moreover, the combination of ESDC can more effectively break the degeneracies between parameters, thus the constraint on EoS of dark energy could be improved significantly: $\sigma(w)= 0.013$.

From the left panel of figure \ref{Fig_CPL}, we can see that, for the constraints on the CPL model [with $w(z)=w_0+w_a{z}/(1+z)$], similar to the case of $w$CDM, E-ELT, SKA1, and DECIGO could offer better constraints on $\Omega_m$, $w_0$, and $H_0$, respectively, but no one could provide a good constraint on $w_a$. The combination of them could give a tight constraint on $w_0$, $\sigma(w_0)\approx 0.03$, which is much better than the result of {\it Planck} 2018 TT,TE,EE+lowE+lensing+SNe+BAO, $\sigma(w_0)\approx 0.08$ \citep{Aghanim:2018eyx}. For the parameter $w_a$, even the combination of E-ELT, SKA1, and DECIGO cannot offer a useful constraint. However, the interesting thing is that adding CMB they could give an effective constraint on $w_a$, $\sigma(w_a)\approx 0.17$, as shown in the right panel of figure \ref{Fig_CPL}.

For the HDE model, the constrained contours and results from E-ELT, SKA1, DECIGO, ESD, and ESDC are shown in figure \ref{Fig_HDE} and table \ref{tab:full}. From the figure, it could be seen that the situation of the HDE model is very similar to that of the $w$CDM model, considering that they have the same number of free parameters. Firstly, the combination of E-ELT, SKA1, and DECIGO can effectively break the degeneracy between parameters and get very tight constrained results. Secondly, for the measurement of $c$, the CMB alone gives a rather weak constraint. Finally, the degeneracy directions of CMB and ESD in all the parameters plans are roughly orthogonal, so that combining them yields more improved results, $\sigma(\Omega_m)\approx 0.0037$, $\sigma(H_0)\approx 0.25$ and $\sigma(c)\approx 0.017$.

Finally, the constrained results of the I$\Lambda$CDM model are shown in figure \ref{Fig_ILCDM}. Although the ability of these three directly measured $H(z)$ data set individually to constrain I$\Lambda$CDM is weak, the combination of them can very effectively break the parameter degeneracies and thus has a strong constraint ability with $\sigma(\Omega_m)\approx 0.01$, $\sigma(H_0)\approx 0.86$ and $\sigma(\beta)\approx 0.071$. From the right panel of figure \ref{Fig_ILCDM}, for CMB, the parameters of I$\Lambda$CDM have too strong degeneracy to be effectively constrained, but ESDC could give a good constraint on the I$\Lambda$CDM. Especially for the coupling parameter $\beta$, its constraint result can be improved to $\sigma(\beta)\approx 0.0031$.

\section{Conclusion}

The direct measurements on $H(z)$ are very important in cosmology because they are more directly related to the EoS of dark energy, $w(z)$, compared with the cosmological distances. However, the $H(z)$ measurements are rather difficult in the traditional ways, no matter in the aspects of redshift range and measurement precisions. Therefore, it is crucial to develop new cosmological probes in the future.
In the forthcoming two decades, multi-messenger and multi-wavelength synergic observations could be one of the key features of the new astronomical era, which would play an important role in helping reveal the nature of dark energy.  

There are several promising ways of directly measuring $H(z)$ in the next 10-20 years. The synergy between redshift drift observations (in different wavebands) and GW standard siren observations will provide the $H(z)$ measurements covering the redshift range of $0<z<5$. Specifically, E-ELT and SKA1-mid will offer the $H(z)$ measurements in the redshift ranges of $2<z<5$ and $0<z<0.3$, using the redshift drift observations in the optical and radio bands, respectively; DECIGO will offer the $H(z)$ measurements in the redshift ranges of $0<z<3$, using the observation for the dipole anisotropy of luminosity distance from GW standard sirens. Thus, in this work, we put forward the multi-messenger and multi-wavelength observational strategy to measure $H(z)$ based on the three projects, and we wish to see whether the future $H(z)$ measurements could greatly impact the dark-energy research. To illustrate this point, we consider five typical dark energy models including $\Lambda$CDM, $w$CDM, CPL, HDE, and I$\Lambda$CDM models.

We find that E-ELT, SKA1, and DECIGO are highly complementary in constraining dark energy models using the $H(z)$ data. Specifically, it is found that the degeneracy directions of them in almost all the parameters plans are rather different, and thus the combination of them could effectively break the cosmological parameter degeneracies. For example, we find that the joint data analysis gives $\sigma(w)\approx 0.02$ in the $w$CDM model and $\sigma(w_0)\approx 0.03$ in the CPL model, which are better than the results of {\it Planck} 2018 TT,TE,EE+lowE+lensing+SNe+BAO.

In addition, we also investigate what effect the combination of these three promising $H(z)$ measurements and CMB will have on breaking the degeneracies between cosmological parameters and what degree of constraint precision can be improved to dark energy. In all dark energy models we consider, the results are similar, i.e., it is difficult to constrain dark energy effectively using the CMB alone, but when these three directly measured $H(z)$ data are combined, the degeneracies between cosmological parameters can be broken significantly to precisely constrain dark energy. All these results strongly suggest that these three promising probes of the late universe can be useful as a complement to the CMB of the early universe.

In the forthcoming 10-20 years, the three projects we are discussing will be completed or launched. E-ELT is currently under construction and is expected to be completed by 2027\footnote{\url{https://elt.eso.org/about/timeline/}}. After a 30-year observation for the Lyman-$\alpha$ absorption lines of the distant quasar, it will be about 2060 estimated optimistically. Currently, SKA Phase 1 is under construction and is planned to be completed in 2028\footnote{\url{https://www.skatelescope.org/}}. Certainly, SKA Phase 2 is also planned for construction, but in this paper, we do not attempt to make detailed forecasts for SKA2 since its precise configuration is yet to be decided. For the detailed discussions of the redshift drift from SKA2 in cosmology, one can refer to Ref. \citep{Liu:2019asq}. After a 40-year observation of SKA1 for redshift drift, it will be about 2070. The proposal of DECIGO and its scientific pathfinder B-DECIGO have been approved, and B-DICIGO is planned to launch in the 2030s to demonstrate the technologies required for DECIGO \citep{kawamura2021current}. DECIGO is optimally expected to obtain a 10-year observation of GW standard sirens before E-ELT and SKA1 complete the observations described above. Although our forecasts are far out, the results of this paper present the implications of these observations for cosmology in the future and the limits of constraints on cosmological parameters. 

\acknowledgments
This work was supported by the National Natural Science Foundation of China (Grants Nos. 11975072, 11875102, 11835009, and 11690021), the Liaoning Revitalization Talents Program (Grant No. XLYC1905011), the Fundamental Research Funds for the Central Universities (Grants Nos. N2005030 and N2105014), the National Program for Support of Top-Notch Young Professionals (Grant No. W02070050), and the National 111 Project of China (Grant No. B16009).

\bibliography{Drift_GW}
\bibliographystyle{JHEP}
\end{document}